\documentstyle[epsfig]{aipproc2}
\newcommand{\del}{\partial}

\newcommand{\x}{{\mathbf{x}}}
\newcommand{\y}{{\mathbf{y}}}

\newcommand{\n}{{\mathbf{n}}}

\newcommand{\f}{\frac}

\newcommand{\bb}{\bibitem}
\begin{document}
\begin{flushright}
YITP-99-16
\\
astro-ph/9903446
\\
March 1999
\end{flushright}
 
\title{CMB Anisotropy in\\ Compact Hyperbolic Universes}

\author{Kaiki Taro Inoue}
\address{Yukawa Institute for Theoretical Physics
\\Kyoto University, Kyoto 606-8502, Japan}
\maketitle

\begin{abstract}
Measurements of CMB anisotropy are ideal experiments 
for discovering the non-trivial global topology of the universe. 
To evaluate the CMB anisotropy in multiply-connected compact
cosmological models, one needs to compute eigenmodes of the
Laplace-Beltrami operator. 
We numerically obtain the eigenmodes on a compact 3-hyperbolic space
cataloged as $m003(-2,3)$ in SnapPea \footnote{SnapPea is a computer program by
Jeff Weeks for creating and studying CH spaces\cite{SnapPea}.} 
using the direct boundary element method, which
enables one to simulate the CMB in multiply-connected compact models
with high precision. The angular power spectra $C_{l}$'s ($2\! \leq\! 
l\!\leq\!18$)
are calculated using computed eigenmodes for $5.4\!\leq\!k<\!10$ and 
Gaussian random approximation for the expansion coefficients  
for $10\!\leq\!k\!<\!20$. Assuming that the initial power
spectrum is the Harrison-Zeldovich spectrum,
the computed $C_l$'s 
are consistent with the COBE data for
$0.1\! \leq\!\Omega_o\! \leq\! 0.6$. 
 In low $\Omega_o$ models, the large-angular fluctuations can be produced at
periods after the last scattering as the 
curvature perturbations decay in the curvature dominant era.
\end{abstract}

\section*{Introduction}
The Einstein equation does not specify the global topology of 
the universe; therefore, there is no a priori reason to believe 
that the space-like hypersurface of the universe is 
simply-connected. If the space-like hypersurface is multiply-connected 
on the scale of the horizon or less, there is a
possibility of discovering the multiply-connectedness by the future 
astronomical observations. 
\\
\indent
In recent years, there has been a great interest in properties of CMB
anisotropy in multiply-connected cosmological 
models \cite{Ste,deO2,Horn,Flat,Circles,Circles2,Weeks}.
Precise measurements of the CMB anisotropy by the future satellite missions  
such as MAP and PLANCK may enable us to find the
fingerprint of the multiply-connectedness in the CMB. Therefore, 
it is very important for us to simulate the CMB anisotropy in 
multiply-connected FRW models. 
\\
\indent
Constraints on the topological identification scales using the COBE
data have been obtained for
some flat models with no cosmological constant \cite{Ste,Flat}.
The large-angular temperature 
fluctuations discovered by the COBE constrain the  
possible number of the copies of the fundamental domain inside 
the last scattering surface to less than $\!\sim$8 for 
these multiply-connected models.
\\
\indent
The mode functions for flat models can be analytically
obtained; therefore, the angular power spectra are obtained straightforwardly.
On the other hand, no closed
analytic expression of the eigenmodes is known for compact
hyperbolic (CH) spaces. Therefore the analysis of the CMB anisotropy 
in CH models has been considered to be quite difficult.
To overcome the difficulty, the author proposed a numerical approach called 
the direct boundary element method (DBEM) for computing eigenmodes of the
Laplace-Beltrami operator \cite{Inoue}. 14 eigenmodes have been computed for 
a ''small'' CH space 
m003(-2,3) with volume $0.98139$ in the SnapPea catalog and 
it is numerically found that the expansion coefficients behave as 
if they are random Gaussian numbers. 
\\
\indent
We briefly describe the DBEM and the 
properties of the expansion 
coefficients which are used for computing the angular 
power spectra for low $\Omega_o$ CH cosmological 
models. We calculate the angular 
power spectra for m003(-2,3) using computed eigenmodes  with small
$k$, and an
approximate method for eigenmodes with large $k$, 
which are compared with the COBE data.
\section*{Numerical Computation of Eigenmodes}
\vspace{0.5cm}
\indent
The advantage of the DBEM is that it enables one to compute the 
eigenfunctions much 
precisely than other methods as it does not rely on the variational
principle and it uses an analytical fundamental solution, namely the free 
Green's function.
\\
\indent   
Let us first consider the Helmholtz 
equation on a compact connected and simply-connected 
M-dimensional domain $\Omega$ in a simply-connected M-dimensional 
Riemannian manifold $\cal{M}$ 
with appropriate periodic boundary conditions on the boundary 
$\del \Omega$,
\begin{equation}
(\nabla^2+k^2)u(\x)=0,\label{eq:helmholtz}
\end{equation}
where $\nabla^2\equiv\nabla^i \nabla_i,~
(i=1,2,\cdot\cdot\cdot,M)$, and $\nabla_i$ is the covariant derivative 
operator defined on $\cal{M}$. A square-integrable function $u$ is the
solution of the Helmholtz equation if and only if 
\begin{equation}
{\cal R}[u(\x),v(\x)]\equiv \Bigl\langle(\nabla^2+k^2)\
u(\x),v(\x)\Bigr\rangle=0,\label{eq:residue}
\end{equation}
where $v$ is an arbitrary square-integrable function called 
\textit{weighted function} and $\langle\,\rangle$ is defined as 
\begin{equation}
\langle a,b \rangle\equiv \int_{\Omega}  ab\, dV. 
\end{equation}      
Then we put $u(\x)$ into the form
\begin{equation}
u=\sum_{j=1}^M u_j \phi_j,
\end{equation}
where $\phi_j$'s are linearly independent
square-integrable functions.  The best approximate solution can be
obtained by minimizing the residue function $\cal R$ for a fixed
weighted function $v(\x)$ by changing the coefficients $u_j$. 
Choosing the fundamental solution
$G_E(\x,\y)$ as the weighted function, Eq.(\ref{eq:residue}) is 
transformed into a boundary integral
equation, 
\begin{equation}
c(\y) u(\y)
+\int_{\del\Omega} G_E(\x,\y) \f{\del u}{\del x^i} n^i \,\sqrt{g}\, dS
-\int_{\del\Omega} \f{\del G_E(\x,\y)}{\del x^i}n^i\, u \,\sqrt{g}\, dS=0,
\label{eq:bem}
\end{equation} 
where $c(\y)\!=\!1$ for $\y \!\in\! \Omega$ and 
$c(\y)\!=\!\f{1}{2}$ for $\y \!\in \! \del \Omega$.
Since $\Omega$ is compact, the discrete eigenvalues are represented as
$k_0\!=\!0<k_1<k_2,\ldots$. The mode $k\!=\!k_1$ is the most 
important one 
that has the longest ''wavelength''defined as $2 \pi/k$. The author
succeeded in computing 14 eigenmodes for $k\!<\!10$ on m003(-2,3)
\cite{Inoue}. $k_1$ is numerically found to be 5.4.
\\
\\
\indent
From now on, we limit our consideration to CH spaces whose universal 
covering space is ${\cal{H}}^3$. We set the curvature radius of
${\cal{H}}^3$ to 1
without loss of generality. 
For convenience, we expand the eigenmodes $u_\nu$,
($\nu\!\equiv\!\sqrt{k^2-1}$) on the CH space in terms 
of eigenmodes $X_{\nu l}Y_{l m}$'s on ${\cal H}^3$,
\begin{equation}
u_\nu=\sum_{l m} \xi_{\nu l m}\,X_{\nu l}(\chi) Y_{l m}(\theta,\phi),
\label{eq:ux}
\end{equation} 
where $X_{\nu l}$ is a radial eigenfunction,
$Y_{l m}$ is a spherical harmonic and $\xi_{\nu l m}$ is an 
expansion coefficient. It has been numerically found that $\xi_{\nu l
m}$'s (for $l\!<\!19$ and $\nu\!<\!9.94$) behave as if they are 
random Gaussian numbers for m003(-2,3)
, which is consistent with the prediction by random matrix theory.
Random Gaussian behavior is also confirmed in a 2-dimensional
CH space \cite{Aur1}.  
Note that some properties of a quantum system whose classical
counterpart is a chaotic system can be explained by
random matrix theory \cite{Bohigas}.
It has also been numerically found that the variance of $\xi_{\nu l m}$'s is
proportional to $\nu^{-2}$. Using these properties, one can
compute the approximate contribution from ''highly-excited'' modes
($k\!>\!>\!1$) to the angular power spectrum.  
\clearpage
\section*{Temperature Correlation}
\begin{figure}
\centerline{\epsfig{file=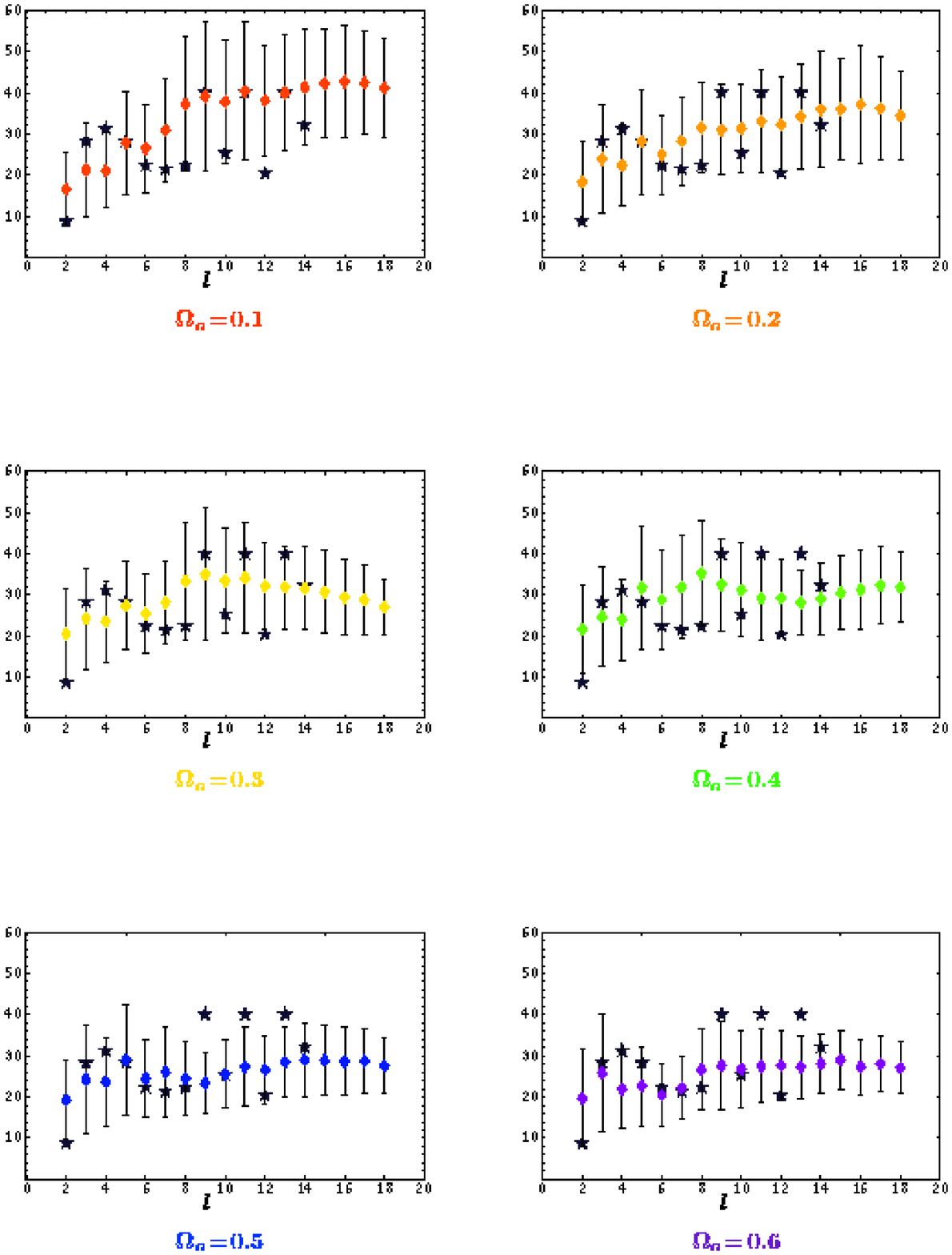}}
\caption{The diamonds show $\delta T/T_l$ 
in $\mu K$ for m003(-2,3) models with $0.1\!\leq\!
\Omega_o\!\leq\!0.6$. The stars show the COBE data
analyzed by Gorski(1996). The 1-$\sigma$ error bars 
are obtained from Monte-Carlo simulation with $N\!=\!10000$
realizations.}  
\label{PS}
\end{figure}

\vspace{0.5cm}
Temperature fluctuations in the multiply-connected FRW cosmological models
can be written as linear combinations (using $\xi_{\nu l m}$ in the 
last section) of independent components of temperature correlations 
in the simply-connected FRW cosmological models.
Assuming that the perturbations are adiabatic and super-horizon scalar type
and the initial fluctuations are random Gaussian, the two-point
temperature correlation in a CH cosmological model is written as
\begin{eqnarray}
\Biggl\langle \f{\delta T}{T}(\n) \f{\delta T}{T}(\n')\Biggr\rangle
\!&=&\!\!\sum_{l,m,\,l',m'} \bigl\langle a_{lm} a^*_{l'm'} \bigr\rangle
Y_{lm}(\n) Y^*_{l'm'}(\n')
\nonumber
\\
\!&=&\!\!
\sum_{\nu,\nu'\!,l,m,l',m'}\!\!\!\!\bigl\langle \Phi_{\nu} \Phi_{\nu'} \bigl
\rangle
~\xi_{\nu l m}\xi^\ast_{\nu' l' m'} 
L_{\nu lm}L^\ast_{\nu' l'm'}\,,
\end{eqnarray}
where
\begin{eqnarray*}
\bigl\langle \Phi_{\nu} \Phi_{\nu'} \bigl
\rangle
\!\!&=&\!\!\f{4 \pi^4~{\cal P}_\Phi(\nu) }
{\nu(\nu^2\!+\!1)\textrm{Vol}(\Omega)}\delta_{\nu \nu'},
\nonumber
\\
\\
L_{\nu lm}(\eta_o,\n) 
\!&\equiv&\! - Y_{lm}(\n) F_{\nu l}(\eta_o),
\nonumber
\\
\\
 F_{\nu l}(\eta_o)
\!\!&\equiv&\!\!\f{1}{3}
\Phi_t(\eta_\ast) X_{\nu l}(\eta_o\!-\!\eta_\ast)
\!+\!\! 2 \!\!\int_{\eta_\ast}^
{\eta_o}\!\!\!\!\!\!d \eta\, 
\f{d\Phi_t}{d \eta}X_{\nu l}(\eta_o\!-\!\eta),
\nonumber
\label{eq:cor} 
\\
\\
\Phi_t(\eta)
\!\!&=&\!\!
\f{5(\sinh^2 \eta-3 \eta\sinh\eta+4 \cosh\eta-4)}
{(\cosh\eta-1)^3}.
\end{eqnarray*}
Here, ${\cal P}_\Phi(\nu) $ is the initial power spectrum, 
$\textrm{Vol}(\Omega)$ denotes the volume of the CH space and 
$\Phi_\nu$ and $\Phi_t(\eta)$ are the $\nu$-component and the time evolution 
of the Newtonian curvature
perturbation, respectively. $\eta_\ast$ is the conformal time of the
last scattering and $\eta_o$ is the present conformal time. 
The diagonal elements ($l\!=\!l'$ and $m\!=\!m'$) give the 
approximate angular power spectrum $\tilde{C}_l$ as
\begin{eqnarray}
(2\,l+1)\,\tilde{C}_l
&=&\sum_{m=-l}^{l} \langle~ |a_{lm}|^2 \rangle
\nonumber
\\
&=&\sum_{\nu,m}\f{4 \pi^4~{\cal P}_\Phi(\nu) }
{\nu(\nu^2+1)\textrm{Vol}(\Omega)}~|\xi_{\nu l m}|^2 |F_{\nu l}|^2 .
\end{eqnarray}
It should be noted that the non-diagonal terms (either $l\!\neq\!l'$
or $m\!\neq\!m'$) are not negligible for large angular (small l)
fluctuations in anisotropic models such as CH models. Therefore,
constrains on the models by using only the angular power spectra $C_l$
are not sufficient. On the other hand, this
property can be considered as the ''fingerprint'' of the multiply-connectedness of the
spatial geometry of the universe.
\\
\indent
Computation of highly-excited eigenmodes is still a difficult 
task since the number of the eigenmodes increases as $k^3$ and the
number of the boundary elements increases as $k^2$.
In order to avoid these difficulties, we assume that $\xi_{\nu l m}$ 's are 
random Gaussian numbers and
the variance is proportional to $\nu^{-2}$. Weyl's asymptotic formula 
gives the approximate values of highly-excited eigenmodes
\begin{equation}
\tilde{\nu}(N)=\Biggl(\f{6 \pi^2 N} {\textrm{Vol} (\Omega)}\Biggr)^{1/3},
\end{equation} 
where $N$ is an integer.
We use pseudo-Gaussian random numbers $\xi_{\nu l m}$ 
that are derived from 14 eigenmodes on m003(-2,3)
using the DBEM for $k\!<\!10$, and random Gaussian numbers 
$\xi_{\tilde{\nu}(N) l m}$ whose variance is 
proportional to $\nu^{-2}$ for $10\!\leq\!k\!<\!20$. 
\\
\indent
In FIGURE \ref{PS}, $\delta T/T_l\!\equiv\![l(l+1)\tilde{C}_l/2
\pi]^{1/2}$  is (diamonds) plotted with the COBE data analyzed 
by Gorski (stars) \cite{Gorski} assuming  that 
the initial power spectrum is the (extended) 
Harrison-Zeldovich spectrum ${\cal{P}}_{\Phi}(\nu)\!=\!Const$. The 1-$\sigma$
error bars are obtained by Monte-Carlo simulation with 10000 realizations. 
$\delta T/T_l$  is 
almost constant in the limit $\Omega_o$ to $1$ for the 
Harrison-Zeldovich spectrum. We see from these figures that $\delta T/T_l$ 
is almost flat for $0.2\!\leq\!\Omega_o\!\leq\!0.6$. Suppression of 
the large angular power due to the long ''wavelength'' cutoff  is
quite mild compared with some flat multiply-connected models since the bulk of the
large angular power comes from the decay of curvature perturbations
well after the last scattering time, which is known as the
integral Sachs-Wolfe effect. Considering the cosmic variance, the
suppression of the large angular power for the $\Omega\!=\!0.1$ model
is still within the acceptable range. 
\section*{Summary}
\indent 
We numerically obtain 14 eigenmodes on a compact hyperbolic (CH) space 
m003(-2,3) with volume $0.98139$ using the direct boundary element
method (DBEM). The temperature fluctuations are written in terms of
the expansion coefficients $\xi_{\nu l m}$ and eigenmodes on the 
universal covering space. For the 14 eigenmodes, $\xi_{\nu l m}$'s
are numerically found to be pseudo-random Gaussian numbers with
variance proportional to $\nu^{-2}$. 
\\
\indent
The angular power spectra are computed using the 14 eigenmodes and an
approximate method for eigenmodes with large k 
which is based on the assumption that the expansion coefficients are 
Gaussian random numbers. In contrast to multiply-connected flat 
models, the suppression of the large angular power is found to be 
so weak that the obtained powers are consistent with the COBE data for 
$0.1\!\leq\!\Omega_o\!\leq\!0.6$. Assuming that the initial perturbations
are adiabatic, constraints on CH models are not so severe as long as one 
uses only the angular power spectra which contain
only isotropic information. 
\\
\indent
However, one must also consider the anisotropic information of the temperature
fluctuations. Contribution of the non-diagonal elements to the
two-point temperature fluctuations is one of the key issues. 
Recently, Bond \textit{et al} have 
obtained much severe constraints on the size of the
topological identification scale for CH models using a method of
images\cite{Pogosyan}.
At the moment, the relation between their result and the author's 
result is not clear.   
Various methods for extracting the anisotropic information 
have been suggested such as a search for circles in
the sky \cite{Circles}, or pattern formation \cite{Spots}. The searches for the
multiply-connectedness in the universe have just begun.
\\
\vspace{0.2cm}
\centerline{\bf Acknowledgments}
\\
\\
\indent
I would like to thank Kenji Tomita and Naoshi Sugiyama for their helpful
discussions and continuous encouragements. 
I am supported by JSPS Research Fellowships 
for Young Scientists, and this work is supported partially by 
Grant-in-Aid for Scientific Research Fund (No.9809834).

\end{document}